\documentstyle[pre,aps,multicol,epsfig]{revtex}
\epsfclipon

\begin{document}   

\tightenlines

\title{Late stages of coarsening in model C}
\author{Julien Kockelkoren and Hugues Chat\'e}
\address{CEA --- Service de Physique de l'Etat Condens\'e, 
Centre d'Etudes de Saclay, 91191 Gif-sur-Yvette, France}
\maketitle
\begin{abstract}
We present a comprehensive picture of (non-critical) domain growth 
in model~C systems where a non-conserved scalar order parameter is coupled 
to a conserved concentration field.
For quenches into the region where the ordered and disordered phases coexist,
we confirm earlier partial numerical
results and find a growth exponent $z=3$. For quenches into the
ordered region, we confirm the theoretical prediction $z=2$.  
Finally we discuss the implications of our results for domain growth in the
microcanonical $\phi^4$-model and we offer some criticism of the work
of Somoza and Sagui on the morphology and wetting properties of domains. 
\end{abstract}



\begin{multicols}{2} 

\narrowtext

\section{Introduction}

\begin{figure}
\centerline{
\epsfxsize=65mm
\epsffile{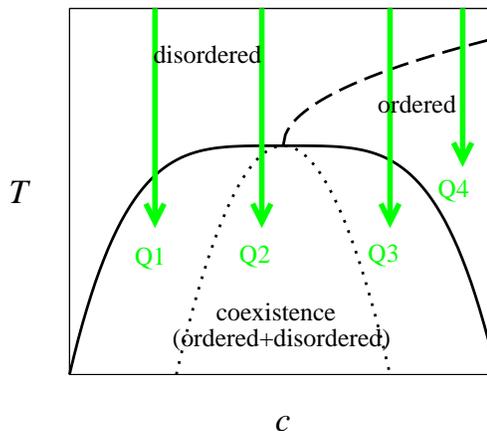}
}
\caption{Typical equilibrium phase-diagram of a model~C in the ($c,T$) plane. 
The solid line is
the coexistence curve, the dashed curve is the order-disorder
transition line. The dotted line
within the coexistence region is the spinodal curve inside
which the three phases are unstable.
Four different types of quenches (Q1, Q2, Q3, and Q4) are shown.
}
\label{f1}
\end{figure}

Coarsening has been a object of intensive study over the last decades
not only because of its experimental relevance but also because it
constitutes what is maybe the simplest case of out-of-equilibrium dynamics. 
For a scalar order parameter, one generally distinguishes 
the non-conserved case, e.g.
the Ising model with Glauber dynamics (model A according to 
the widely-used classification of \cite{HOHA}), from the conserved case,
e.g. the Ising model with Kawasaki dynamics (model B). 
It is now well-established that in these cases phase ordering is 
characterized by a single lengthscale $L$ growing algebraically with time
($L\sim t^{1/z}$) with the growth exponent $z$ taking two different values,
$z=2$ for the non-conserved/model~A case, and $z=3$ for conserved
order parameter systems like model~B \cite{BRAY}.
Here we are interested in the more complicated and therefore less
well understood case of a non-conserved order
parameter coupled to a conserved concentration (so-called model~C). 
Examples of this situation can be found in various physical systems, 
e.g., intermetallic alloys (see \cite{gfl} and references therein),
adsorbed layers on solid substrates \cite{binder82} and
supercooled liquids \cite{tanaka}.
In these systems, the decomposition process (described by the conserved
concentration $c$) and the ordering process (described by the
non-conserved order parameter $\psi$) are coupled. 
Let us consider an $A-B$ alloy on a square lattice
in order to illustrate this. 
A state in which all $A$ atoms are
surrounded by $B$ atoms is energetically favorable. The ordered state
thus consists of two symmetric sublattices, and we can define an order
parameter $\psi$ as half of the difference between the $A$-concentration
in each sublattice.
In this way, $\psi=1$ when
all the $A$ atoms are on the one sublattice and $\psi=-1$ when they
are on the other. 
At high temperature, a disordered state $\psi=0$ arises. 
It is now easy to realize that for asymmetric initial conditions
(i.e. an unequal amount of $A$ and $B$ atoms) the
system will not be able to completely order (strictly speaking, this is 
only true at low-enough temperature).  Hence, as opposed to
model~A, the disordered phase can coexist with the two
ordered phases. On a typical equilibrium phase diagram 
in the concentration-temperature  ($c$-$T$) plane
(Fig.~\ref{f1}), one can thus distinguish, apart from a disordered
region and an ordered region, a coexistence region. 
The dashed line separating the ordered and disordered regions 
marks a second-order
phase transition. In the spinodal region inside the coexistence
region (dotted line), the three phases are thermodynamically unstable. 

Models have been proposed to account for various aspects of the 
morphology and of the kinetics of the experimental systems (see for
instance \cite{gfl} and references therein).  
From the more theoretical point of view of universality issues, 
the situation is not quite satisfactory. 
For instance, the critical exponents, and in particular the dynamic
critical exponent, are still debated \cite{stauffer,BZ-PRL,bsh}. 
A renormalization group analysis turns out to
be more delicate than in the case of model~A \cite{brezin,Oerding}. 
Our goal here is to clarify the a priori simpler problem of
domain growth below criticality, when the system is quenched down from
a high-temperature state.
Notable but partial results, somewhat scattered in the literature,
have been obtained in the past.
For quenches into the spinodal region with droplet morphology 
(quench Q2 of Fig.~\ref{f1}) San Miguel et
al. \cite{SGDS81} have predicted the model~B exponent $z=3$.  
Numerical simulations in the
context of a Oono-Puri ``cell model'' have been
found to be consistent with this prediction\cite{Ohta,Chakrabarti}.
On the other hand, Elder et al. \cite{emgd81} have predicted $z=2$ for
quenches above the tricritical temperature, i.e.\ in the ordered region
(quench Q4). To the best of our knowledge, 
this has not been verified numerically. 

Our goal here is to give a complete picture of (non-critical) domain growth
in model~C, considering, within a single system introduced in 
Section~\ref{s2}, all four possible types 
of quenches illustrated in Fig.~\ref{f1}. This is done in Section~\ref{s3}.
In Section~\ref{s4},  in the sake
of comprehensiveness, we come back to the two following 
unsettled issues discussed recently in works about model~C systems.
The microcanonical $\phi^4$
model \cite{CAIANI2,CAIANI1,BZ-PRL}, is a type of model~C since
the order parameter is coupled to the (conserved) energy.
Zheng has suggested in a recent paper \cite{BZ} that 
domain growth is characterized by a non-trivial value of $z$
($2<z\simeq 2.65 <3$).  A more careful study 
by us showed that the data are in fact consistent with the model A
exponent $z=2$ \cite{bz-comm}. Here we detail to which phase of model~C
the microcanonical $\phi^4$ model belongs.
The morphology of domains and the related ``wetting'' issues have 
also been a point of contention in the past. In experiments, it
has been observed that neighboring ordered domains do not merge
\cite{mik}. A possible explanation proposed in \cite{wk94} is that the 
domains are different variants of the same ordered structure. 
The simulations of \cite{gfl} seem to indicate that ordered domains
do not join but ``stay separated by narrow channels of
the disordered phase'': the antiphase boundaries appear to be wetted
by the disorder phase.  But Somoza and Sagui
\cite{Somoza} have found on the contrary that inside the 
coexistence region the two ordered phases may be in direct contact.   
We revisit their work and resolve the controversy.
A summary of our results is given in Section~\ref{s5}.
  
\section{The model}
\label{s2}

We choose one of the simplest versions of
model C which can be written as follows:  
\begin{eqnarray}
\frac{\partial \psi}{\partial t} & = & -\Gamma_{\psi} \frac{\delta
F}{\delta \psi} + \eta \label{c1}\\
\frac{\partial c}{\partial t} & = & \Gamma_{c} \nabla^2 \frac{\delta
F}{\delta c} + \eta'. \label{c2}
\end{eqnarray}
Here $\Gamma_{\psi}$ and $\Gamma_{c}$ are kinetic coefficients, $\eta$
and $\eta'$ represent thermal noise and 
$F[\psi,c]$ is the Ginzburg-Landau free energy functional 
which takes generally the following form:
\begin{equation}
F[\psi,c] = \!\int\! d^n r \left[ f(c,\psi) + \frac{K_c}{2} (\nabla c)^2 +
\frac{K_{\psi}}{2} (\nabla \psi)^2 \right],
\end{equation}
where $K_c$ and $K_{\psi}$ are diffusion constants.
The function $f(c,\psi)$ should satisfy a few constraints. Firstly
it has to be symmetric in $\psi$: $f(c,\psi)=f(c,-\psi)$. It
should also allow for the coexistence of a disordered phase $(c_n,\psi_n)$
with the ordered phases $(c_s,\pm\psi_s)$ \cite{notation}. These points correspond
to minima in the free energy landscape. 
A possible choice is \cite{Somoza,fd97}
\begin{equation}
f(c,\psi)=\psi^2(1-\psi^2)^2 + \alpha (\psi^2-c)^2 \;,
\end{equation}
where $\alpha$ governs the coupling between the two fields.
The minima are then  $(c_n,\psi_n)=(0,0)$ corresponding to the disordered
phase and $(c_s,\psi_s)=(1,\pm 1)$ corresponding to the ordered phase.

It is instructive to analyze the model in terms of its uniform fixed
points $(c_0,\psi_0)$. They lie on the curves where $f_{\psi}=0$
which are the $c$-axis $\psi_0=0$ and the polynomial $c_0=\frac{1}{4
\alpha} ( 1-3\psi_0^2)(1-\psi_0^2) + \psi_0^2$. 
Their stability is determined by the signs of 
$f_{\psi \psi}$ and $(f_{cc}f_{\psi\psi}-f_{c\psi}^2)$ \cite{SGDS81,fd97}. 

Fixed points on the $c$-axis change stability at the so-called ordering
spinodal $c_1$: they are stable for $c < c_{1} = \frac{1}{4
\alpha}$. The point on the second curve where stability changes is the
conditional spinodal $c_{\sigma}$. Stability occurs when $c > c_{\sigma} =
\frac{4 \alpha - 1}{6 \alpha}$.

\begin{figure}
\centerline{
\epsfxsize=60mm
\epsffile{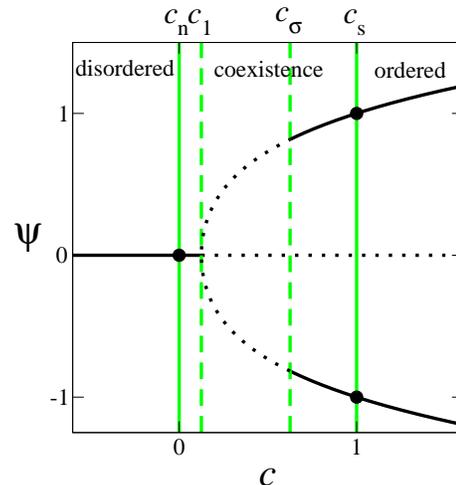}
}
\caption{($c$,$\psi$)-plane. Solid lines: stable uniform fixed
points. Dotted lines: unstable uniform fixed points. 
The minima of $f$ are indicated by filled circles. 
Grey lines delimit the coexistence domain. Dashed grey lines delimit the spinoidal region} 
\label{f2}
\end{figure}

Consider now the evolution of a two-dimensional system starting from 
random initial conditions disordered around $\psi=0$ with some fixed 
concentration $c_i$. We briefly present the expected follow-up of
quenches at various $c_i$ values before describing our corresponding
numerical results in the next section.

\noindent
--- For $c_i < c_n$ (disordered phase), 
the system evolves towards  the stable point $(c_i,0)$.

\noindent
--- For $c_n < c_i < c_s$, the three phases $(0,0)$ and $(1,\pm 1)$ can in
principle coexist. 

\noindent
--- For $c_n < c_i < c_1$ (quench Q1), however, $(c_i,0)$ is still stable 
and it will be reached if initial
fluctuations are small enough. 

\noindent
--- Inside the region $c_1 < c_i < c_{\sigma}$ (quench Q2),
the system necessarily undergoes spinodal decomposition.
For $c_{\sigma} < c_i < c_s$ (quench Q3), 
even though the ordered fixed points are now stable, nothing is changed
since the initial conditions (chosen around $\psi=0$) lead to the formation
of disordered domains.

\noindent
--- For initial concentrations $c_i > c$ (quench Q4), the system will not
phase separate but will order, so we identify this region with the
ordering regime.

\section{Numerical results}
\label{s3}

In order to investigate the late stages of domain growth,
Eqs.~(\ref{c1}-\ref{c2}) are now studied numerically. 
We perform quenches from the high-temperature disordered phase to
$T=0$.
(The noise term in Eqs.~(\ref{c1}-\ref{c2}) can be set equal to zero, 
since the growth exponent is expected to be determined 
by the ``$T=0$ fixed point'' \cite{BRAY}.) 
In practice, initial conditions are randomly
distributed around $(c_i,0)$. More specifically, we choose $\psi \in
[-0.1 : 0.1]$ and $c \in [c_i-0.05:c_i+0.05]$. 
  
Equations (\ref{c1}-\ref{c2}) are numerically
solved on a two-dimensional grid.
For the time integration we use Euler's method and we 
approximate the Laplacian by:
\begin{equation}
\nabla^2 y = \frac{1}{\Delta x^2} \left[ \frac{1}{6} \sum_{{\bf j} \in {\cal V}_i^1} y_{\bf j} +
\frac{1}{12} \sum_{{\bf j} \in {\cal V}_i^2} y_{\bf j}  - y_{\bf i}\right]
\end{equation}
where ${\cal V}_i^1$ and ${\cal V}_i^2$ are the sets of
nearest and next-nearest neighbors of site $i$.

In most regimes studied below, $\Delta x=1.0$ and $\Delta t=0.01$ allow for a 
smooth representation of both fields. Without loss of generality,
we set: $K_c = K_{\psi} =1$ and $\alpha=4$.
We record the typical domain size $L$ of both the order parameter and the
concentration fields, as determined by the
mid-height value of $C_{c,\psi}(r,t)$, the normalized two-point correlation
function calculated for simplicity along the principal axes of the
lattice using the the reduced ``spin'' variables $\sigma_\psi \equiv {\rm
sign}(\psi)$ and  $\sigma_c \equiv {\rm sign}(c-c_i)$:
\begin{equation}
C(r,t)=\langle \frac12 (\sigma_{i+r \, j}(t) + \sigma_{i \,
j+r}) \sigma_{ij}) \rangle - \langle \sigma_{ij} \rangle^2 
\end{equation}

\subsection{Quenches into the coexistence regime: droplet regime}

Quenches within the disordered regime being uninteresting, we first discuss 
Q1 quenches, i.e. quenches into the coexistence
region, but left of the spinodal ($c_n<c<c_1$). 
Here the system possesses a stable
fixed point $(c_i,0)$ but will be able to locally reach the ordered
fixed points if initial fluctuations are large enough. The precise
dependence on initial conditions is in fact a non-trivial problem that
we have not studied systematically. 
This shows that the spinodal lines do not have a strict meaning in the 
presence of fluctuations, a well-known fact (see for
instance \cite{Binder}). 

\begin{figure}
\centerline{
\epsfxsize=8cm
\epsffile{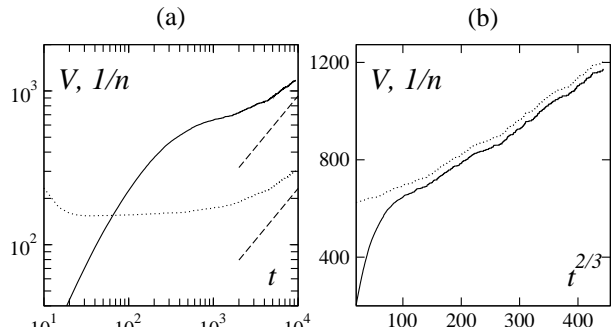}
}

\caption{Domain growth at $c_i=0.1$. (a) $V$ and $1/n$ (dotted lines) vs. $t$
in log-log plot. The dashed lines indicate the expected asymptotic behavior
$V$,$1/n ~ t^{2/3}$ (b) $V$ and $1/n$ vs. $t^{2/3}$ ($1/n$ is multiplied
with some factor so that it can diplayed together with $V$).
System size is $2048 \times 2048$.  
} 
\label{f3}
\end{figure}

\begin{figure}
\centerline{
\epsfxsize=4cm
\epsffile{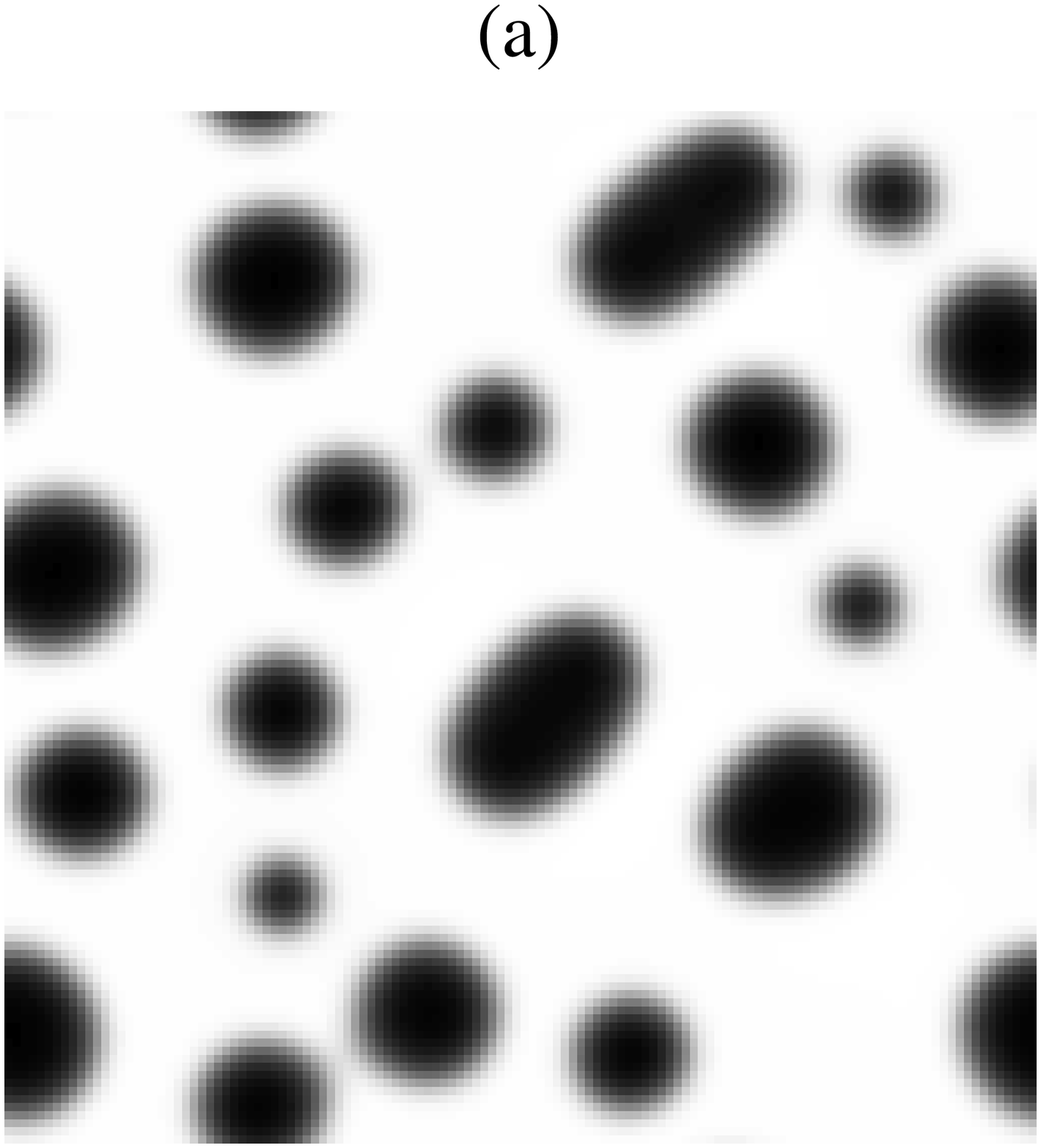}
\epsfxsize=4cm
\epsffile{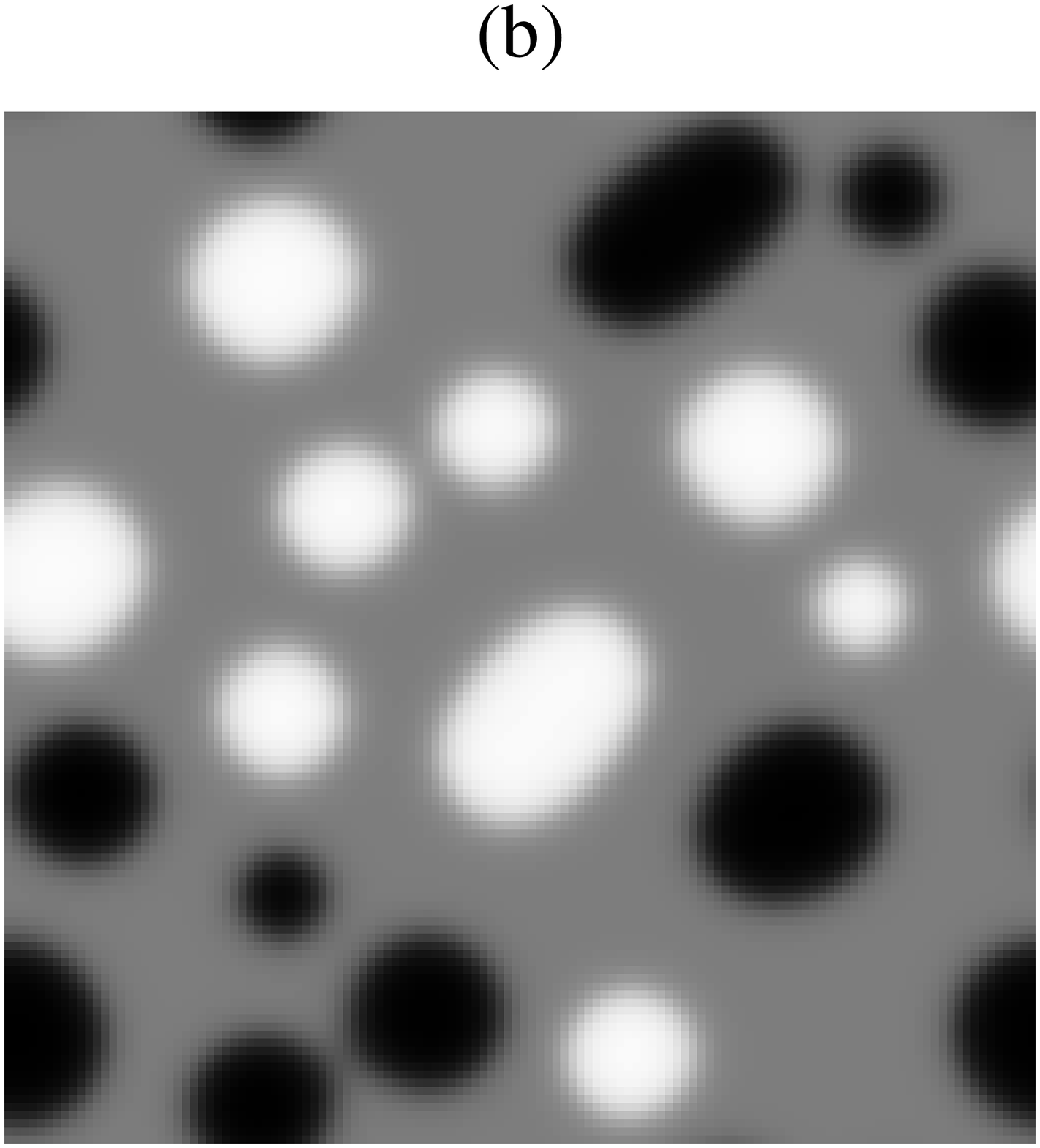}
}

\caption{Snapshots of coarsening in droplet regime ($c_i=0.25$) for a system of
$128 \times 128$ at $t=200$. High values of the field correspond to
dark areas. (a)
concentration field: white corresponds to the disordered phase $c=0$,
black to the ordered phase $c=1$, (b) order parameter field: grey corresponds
to the ordered phases $\psi=0$, white and black correspond to the
ordered phases $\psi=\pm 1$.  
} 
\label{f4}
\end{figure}

When decomposition occurs, the ordered phase is the minority phase, 
and, similarly to what is observed in model~B, 
droplets of either ordered phases form. A similar configuration
is shown in Fig.~\ref{f4}. 
The growth of ordered droplets is constrained by the conservation law. 
The smallest droplet disintegrates to the benefit of the larger ones. 
We thus expect a Lifshitz-Slyozov growth law $L \sim t^{1/3}$. 
In our simulations, this behavior is not easily observed, due
to the presence of a long transient during which the ordered domains
``nucleate'' out of the metastable state. In particular, the growth of the
typical lengthscales $L_{c,\psi}$ does not show the expected scaling
although the trend at large times is good. In this regime, though,
the droplet morphology of the ordred domains can be exploited to
provide a more direct measurement of growth. In Fig.~\ref{f3},
we show the evolution of
the total number of the ordered droplet,s $n$, and of the average
droplet volume $V$. These quantities scale respectively as $t^{-2/3}$ 
and $t^{2/3}$.
Due to a long crossover, the expected behavior is only clearly seen when
plotting $V$ and $1/n$ against $t^{2/3}$ (Fig.\ref{f3}b).

\begin{figure}
\centerline{
\epsfxsize=8cm
\epsffile{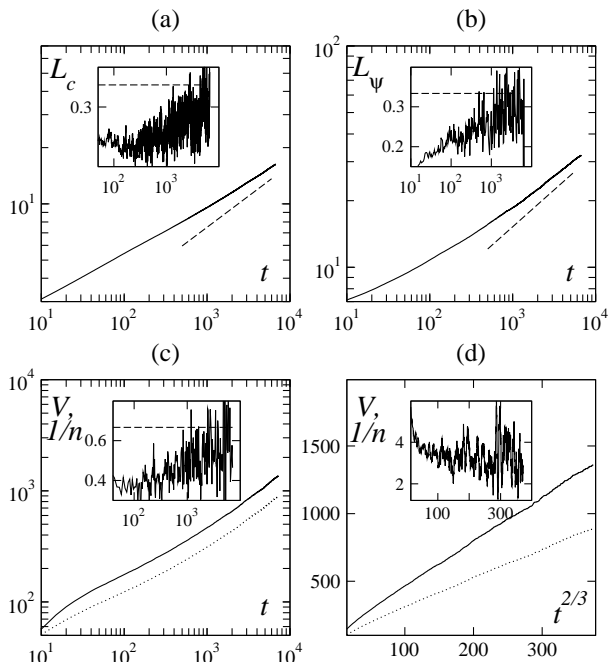}
}

\caption{
Domain growth in the droplet regime ($c_i=0.25$) (a) $L_c$ vs $t$ (b)
$L_{\psi}$ vs $t$ (c) $V$ and $1/n$ (dotted lines) vs $t$ (d)
$V$ and $1/n$ vs $t^{2/3}$. (a),(b) and (c) are on logarithmic scale, (d)
is in linear scale. The dashed lines indicate the expected scaling /
exponent values. 
Insets: local slopes calculated by a running average of the derivated 
signal over a window of fixed size in the corresponding variable 
(i.e. $\log_{10}t$ or $t^{2/3}$ here). System size is $4096 \times 4096$.
} 
\label{f5}
\end{figure}

The same scenario is observed for Q2 quenches provided the ordered
phase be the minority phase, i.e. for $c_1<c_i < \frac{c_n + c_s}{2}=0.5$.   
A typical snapshot is shown in Fig.~\ref{f3}. We have also verified
that indeed a growth exponent $z=3$ is observed. Now the crossover is
shorter and we see that the local slope of $\log L_{c,\psi}$ vs $\log t$
approaches $\frac13$ at late times (see Fig.~\ref{f5}ab). The same
conclusion is reached studying the total number of domains and their
average volume (Fig.~\ref{f5}cd). 

\subsection{Quenches into the coexistence regime: interconnected regime}

Due to the coupling to the non-conserved order parameter, the symmetry
between high and low concentration phases, which exists in model~B, is
broken. For $c_i > 1/2$ the minority phase (now the low
concentration phase) does not form droplets but is instead concentrated
along lines which correspond to interfaces of the
non-conserved field. In the context of alloys one speaks of ``wetting
of antiphase
boundaries''. Looking at the non-conserved field, we first observe
model~A like coarsening which then slows down, 
since the interfacial regions separating ordered domains 
get wider (made of the disordered phase, their volume
is conserved). We thus again encounter the situation
where the ordering is constrained by the conservation law, but with the
difference that the disordered phase has a particular morphology. 
It is therefore not obvious that the growth exponent will be $z=3$. The 
derivation of \cite{SGDS81} is only valid in the case of isolated droplets.
However, the Lifshitz-Slyozov law is known to apply more generally, provided
that 1) coarsening is driven by surface tension, 2) transport is by
diffusion through the bulk and 3) the length scale that describes the
coarsening process is the only relevant length scale in the system
\cite{langer}. 
It is easy to convince oneself that the first two conditions are
fulfilled. At first sight, however, it appears that there are two
lengthscales: one associated with the thickness of the disordered
domains, $d$ (equivalently of the interfaces between ordered domains),
the other with their curvature (equivalent to the lengthscale of the
ordered domains), $L$. One has to realize that the total interface length $l$
is inversely proportional to $L$. Since the volume occupied by
the interface is conserved we also have $d \sim 1/l$, and therefore $d
\sim L$.  One thus expects again a growth exponent $z=3$.

The above discussion not only applies to Q2 quenches 
($1/2<c_i<c_\sigma$), but also to Q3 quenches ($c_\sigma<c_i<c_s$).
A uniform configuration $(c_0,\psi_0)$ is then stable, but for
symmetric initial conditions there are always interfaces where
$\psi=0$ and where it is thus more favorable to have $c<c_i$. In the
``ordered'' domains $c$ will then increase such that we will end up
with all the three phases.  

\begin{figure}
\centerline{
\epsfxsize=4cm
\epsffile{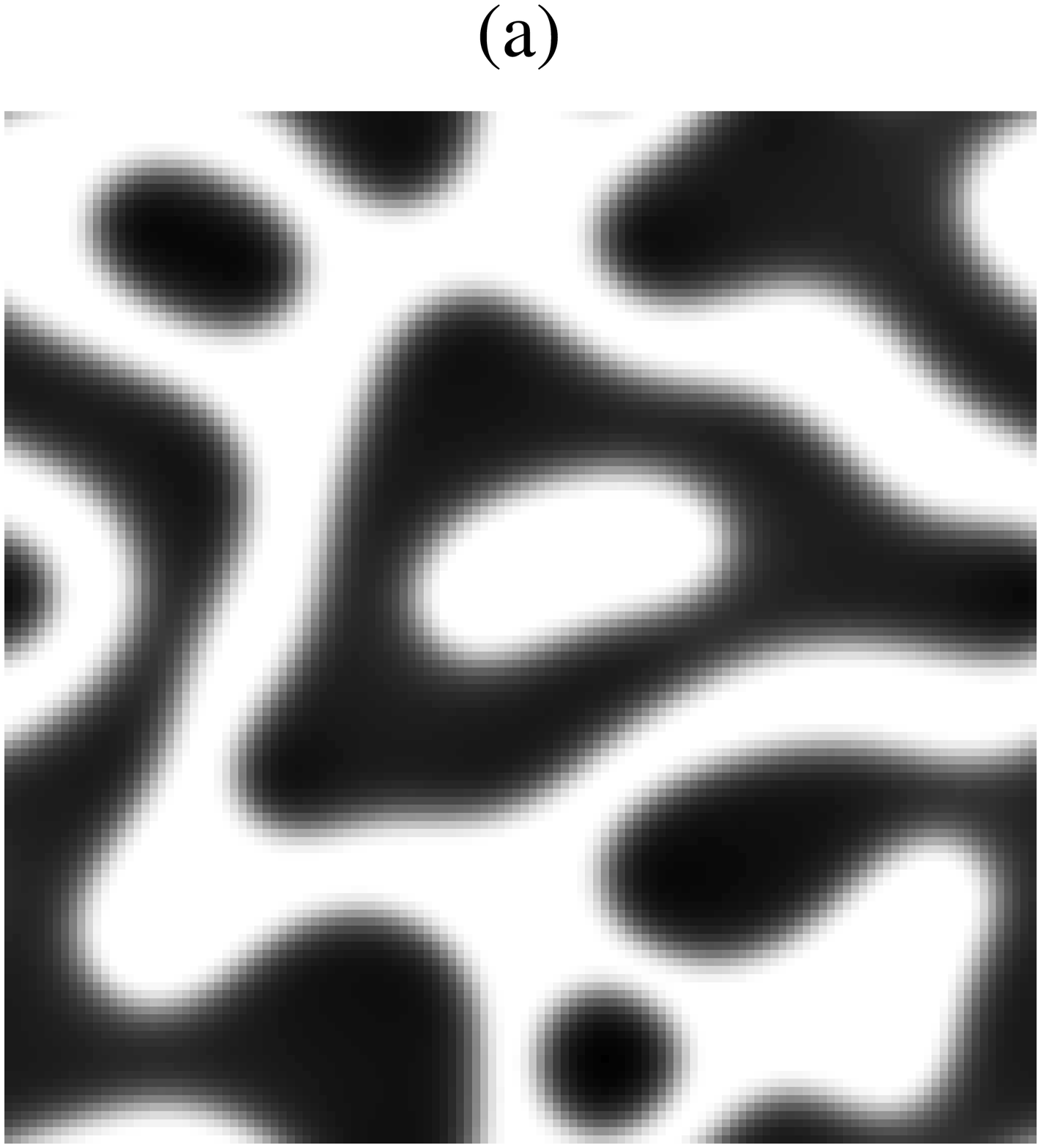}
\epsfxsize=4cm
\epsffile{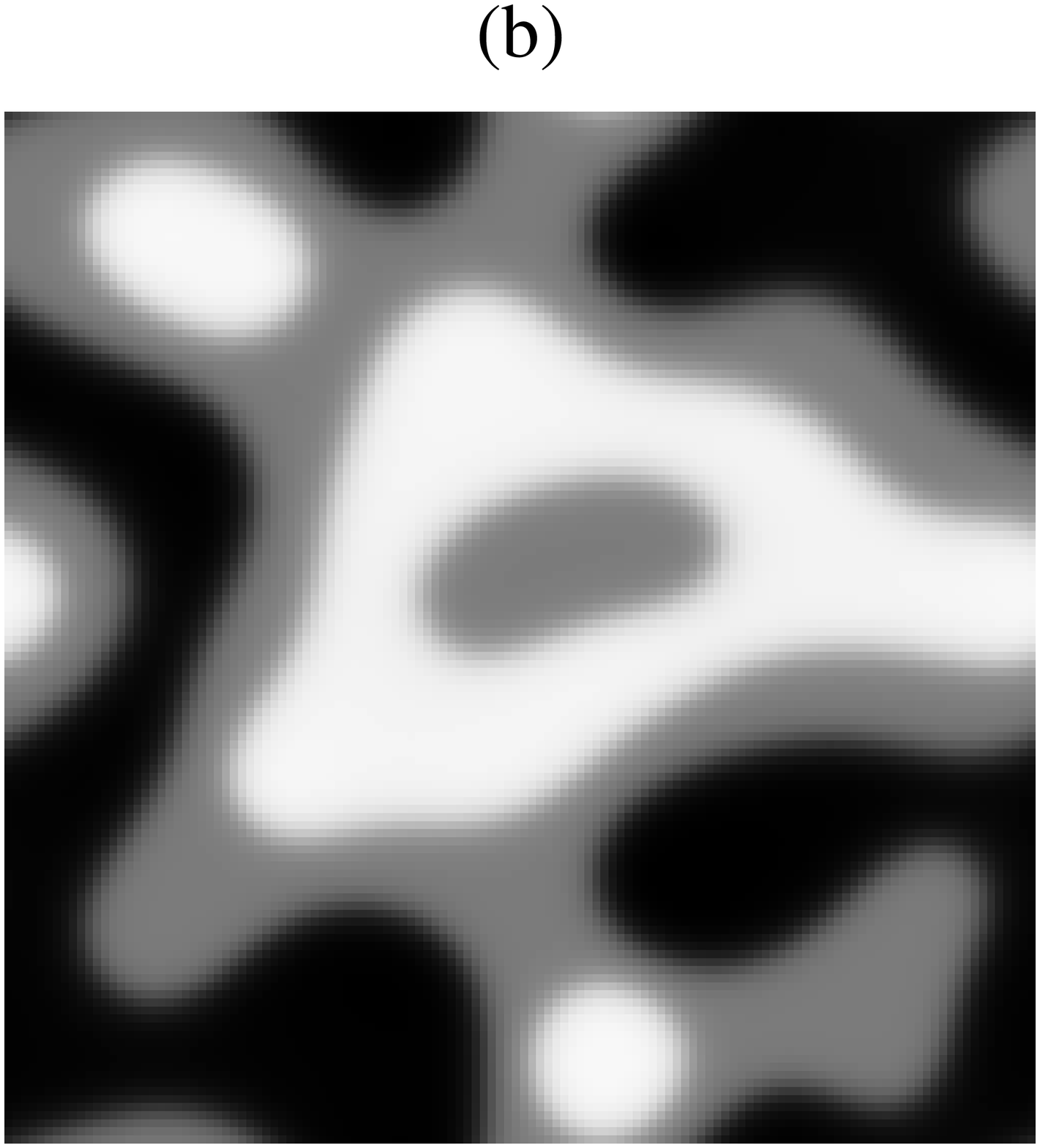}
}

\centerline{
\epsfxsize=4cm
\epsffile{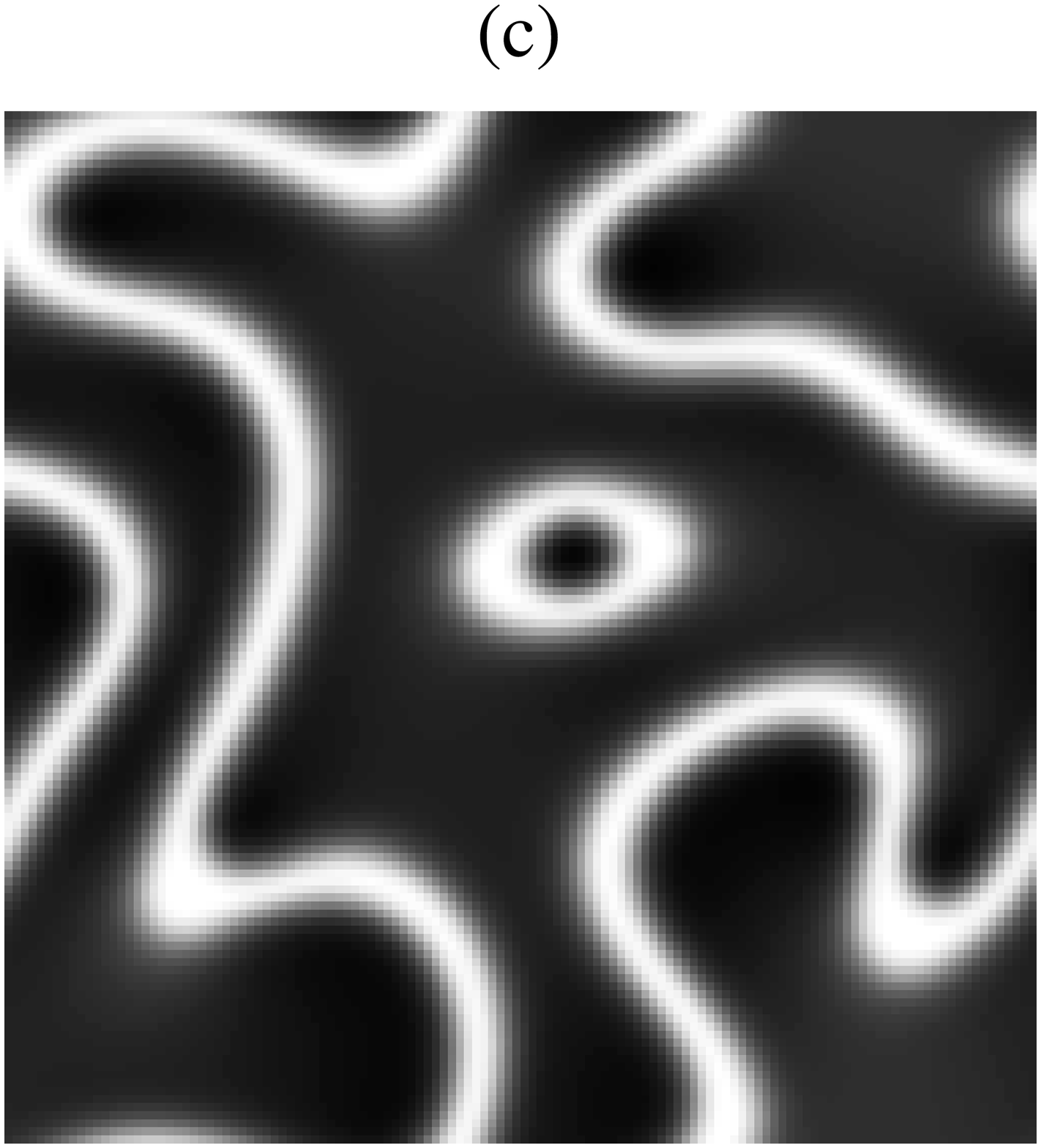}
\epsfxsize=4cm
\epsffile{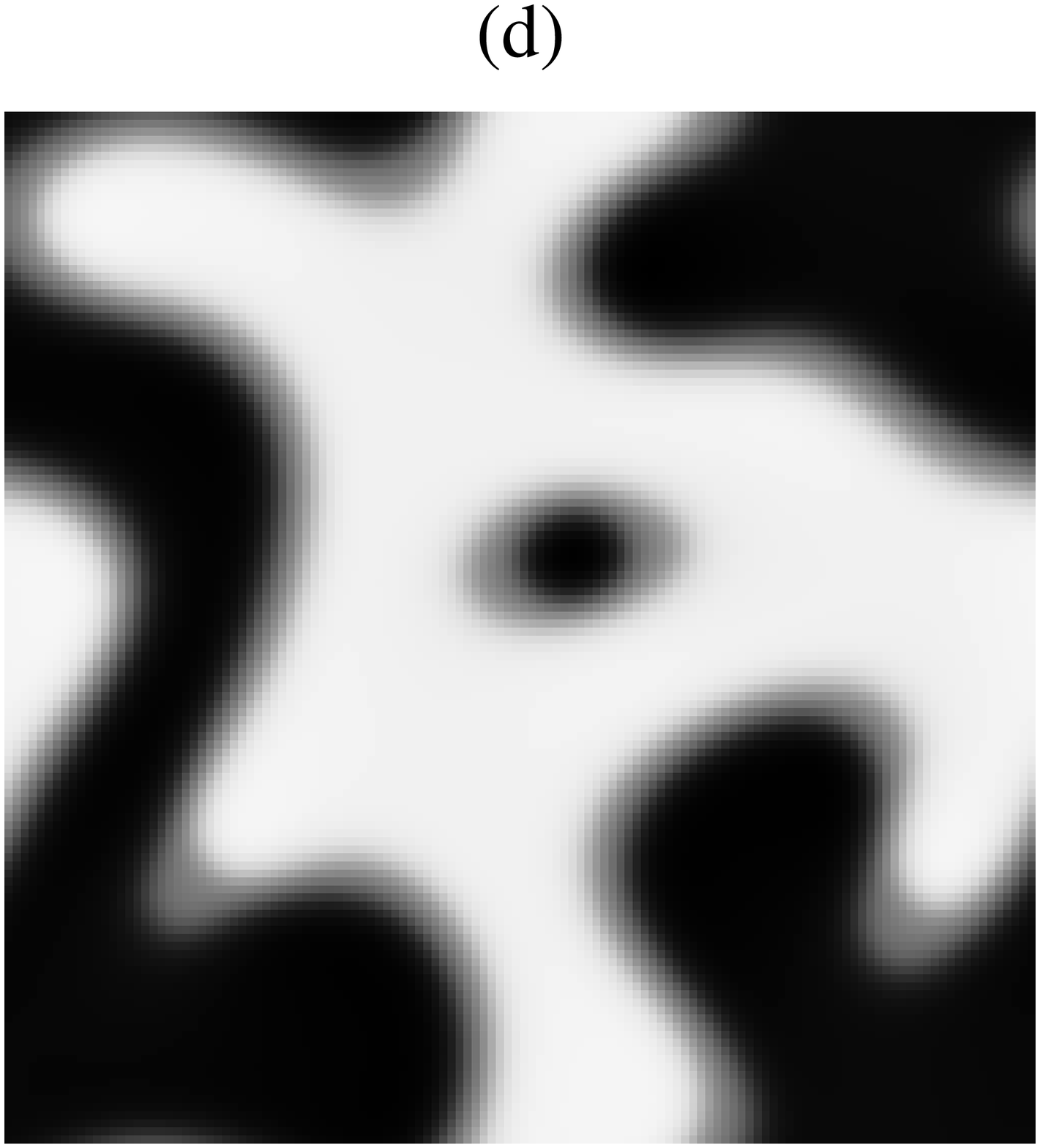}
}
\caption{Snapshots of coarsening in interconnected regime 
(a) Concentration field and (b) order parameter field at $c=0.5$ at $t=200$
(c), (d) id. at $c=0.75$ at $t=120$
System size and greyscales as in Fig.~\ref{f4}.
} 
\label{f6}
\end{figure}

\begin{figure}
\centerline{
\epsfxsize=75mm
\epsffile{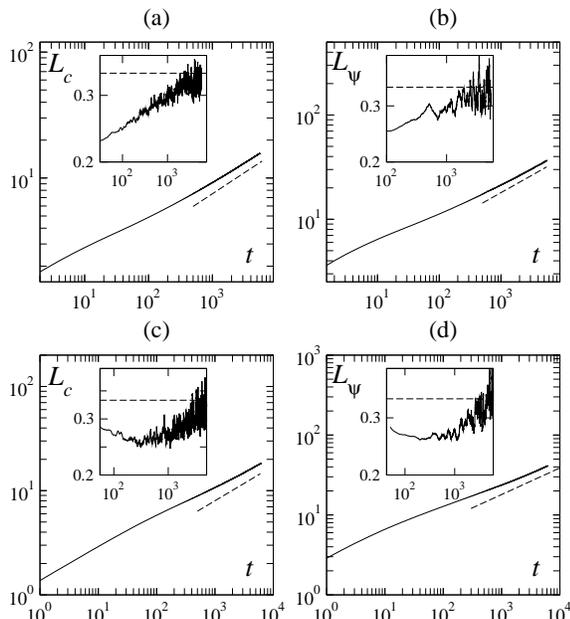}
}

\caption{Domain growth in interconnected regime
(a) $L_c$ vs $t$ ,(b) $L_{\psi}$ vs $t$  both at $c=0.5$, 
(c) $L_c$ vs $t$ ,(d) $L_{\psi}$ vs $t$  both at $c=0.75$
The local slopes in the insets are calculated as explained in Fig.~\ref{f5}. 
System size $4096 \times 4096$. 
} 
\label{f7}
\end{figure}

All this is confirmed by our numerical results. In Fig.~\ref{f6} we show
snapshots of order parameter and concentration at late times for
$c_i=0.5$ and $c_i=0.75$. In Fig.~\ref{f7} we plot, for the same
concentrations, the typical lengthscale of ordered domains and
disordered domains. They behave similarly.  
We can also identify two transient effects. Initially,
rapid growth takes place due to the ordering process.
After that, when the antiphase
bounderies are wetted, the growth will actually be slower than $L \sim
t^{1/3}$. Since the disordered phase has a large surface compared to
its volume, surface diffusion will dominate at short times, and this 
is known to lead to a $L \sim t^{1/4}$ growth law \cite{huse}.

\subsection{Ordered regime}
 
\begin{figure}
\centerline{
\epsfxsize=75mm
\epsffile{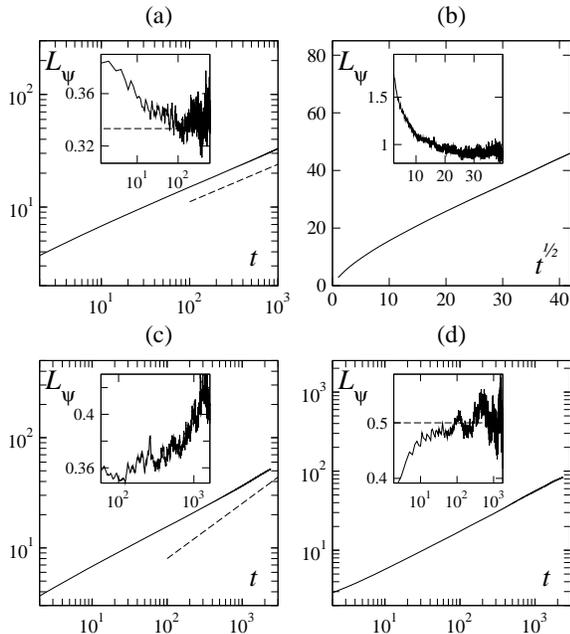}
}

\caption{Growth of typical order parameter domain size $L_{\psi}$ (a)
$c_i=0.95$ (b),(c) $c_i=1.0$ (b): $L$ vs. $t^{1/2}$ (d) $c_i=1.5$.
The local slopes in the insets are calculated as explained in Fig.~\ref{f5}.
The system size is $4096 \times 4096$.
} 
\label{f8}
\end{figure}

For $c_i \geq c_s=1$ (Q4 quenches) no spinodal decomposition occurs. If
initially a domain of the disordered phase exists, it disappears.
Although the concentration is not completely uniform but somewhat
lower at the interfaces of the ordered domains, as seen in
Fig.~\ref{f8}, this does not have any
influence on the coarsening of the order parameter.  Thus, the model~A
exponent $z=2$ should be observed for $c_i \geq 1$. Again this is
confirmed in our simulations. We have measured the typical domain size
for the order parameter around $c_s$. As can be seen in Fig.~\ref{f8} a
growth exponent $z=3$ is still observed for $c_i$ slightly below
$c_s$, at $c_i=0.95$. At $c_i=c_s$, growth is clearly faster but the local
exponent (inset of Fig.~\ref{f8}c) does not reach $\frac{1}{2}$ 
in accessible times. 
The behavior, though, is consistent with
$z=2$. 
This is most clearly observed plotting $L$ againt $\sqrt{t}$ 
(Fig.~\ref{f8}b).
Increasing $c_i$ beyond $c_s$, the crossover is now short enough and
the $z=2$-behavior is apparent even in a log-log plot (Fig.~\ref{f8}d).

\begin{figure}
\centerline{
\epsfxsize=4cm
\epsffile{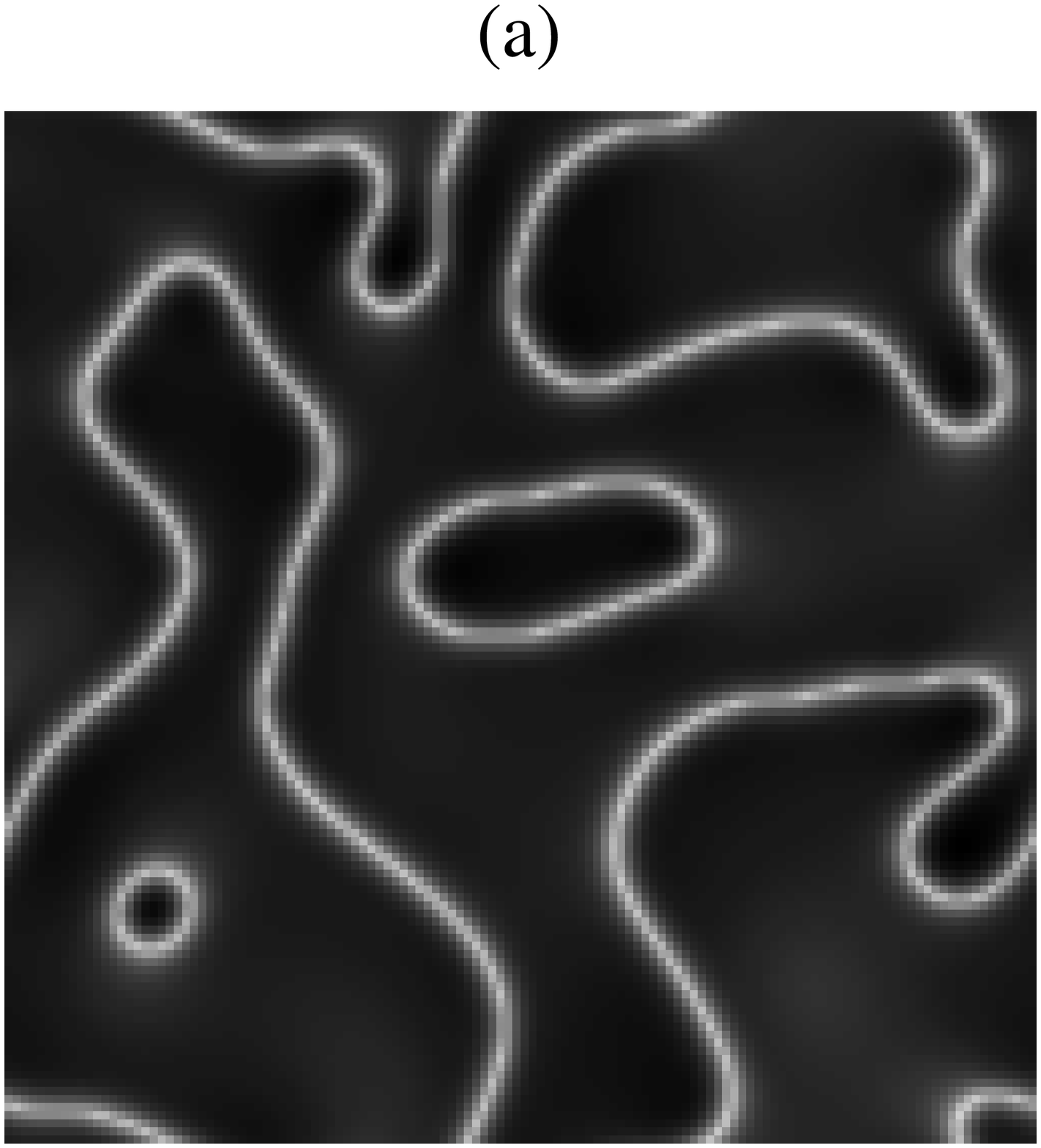}
\epsfxsize=4cm
\epsffile{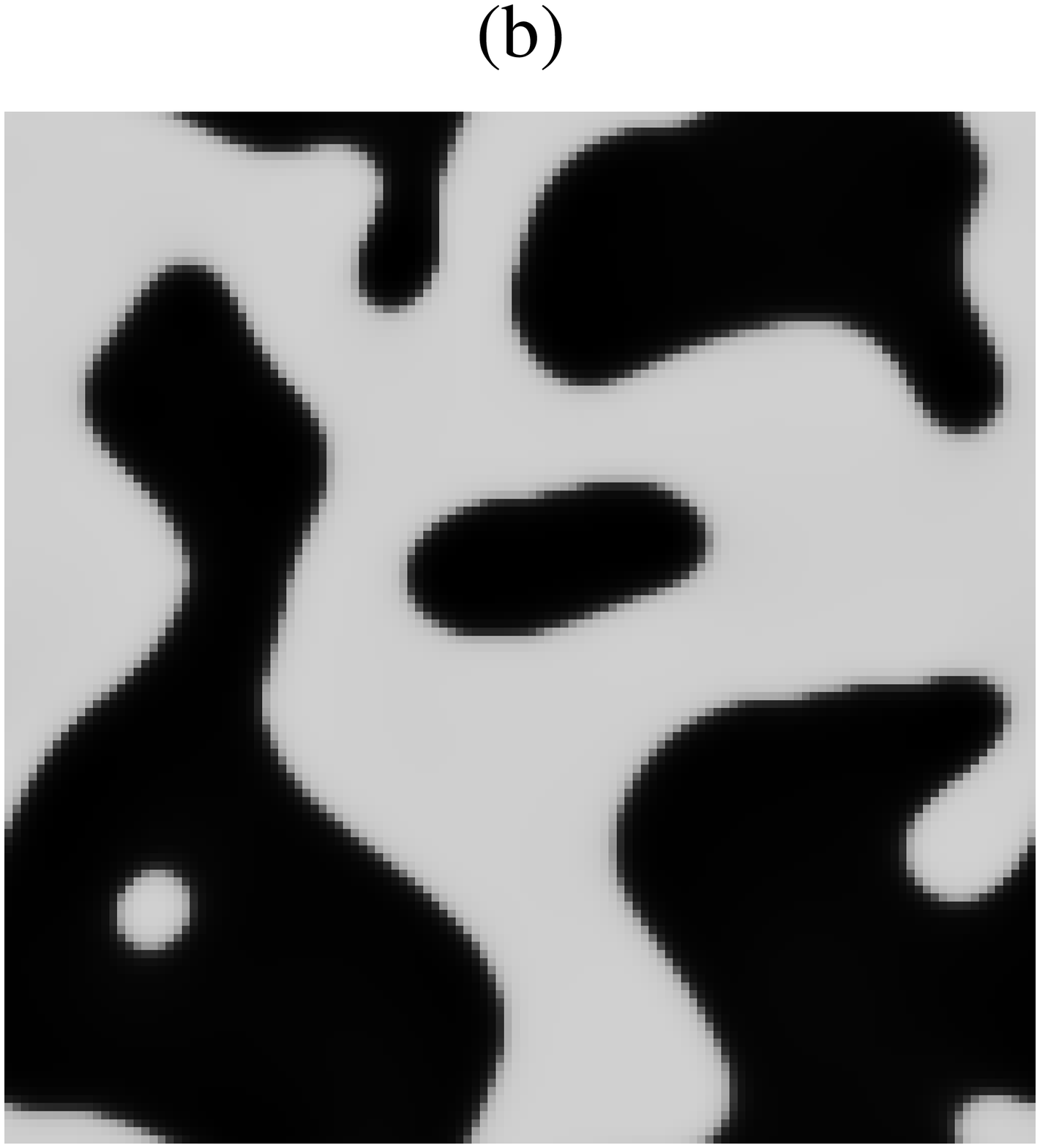}
}

\caption{Snapshots of coarsening in ordered regime ($c_i=1.5$) at
$t=50$ (a) concentration field (b) order parameter field.
System size and greyscale as in Fig.~\ref{f4}.
\label{f8snap}
} 
\end{figure}

\section{Discussion}
\label{s4}

\subsection{The microcanonical $\phi^4$ model}

The microcanonical $\phi^4$ model has lately received renewed
attention largely because it offers an interesting bridge between
statistical mechanics and deterministic dynamics
\cite{CAIANI1,CAIANI2,BZ-PRL}.
Its equations of motion derive from the well known lattice $\phi^4$  
Hamiltonian. They can be written:
$$
\ddot{\phi_i} = \sum_j (\phi_j-\phi_i) + m^2 \phi_i -\frac{g}{6} \phi_i^3
$$
where the sum is over the 4 nearest neighbors of site $i$ on a square lattice.

\begin{figure}
\centerline{
\epsfxsize=4cm
\epsffile{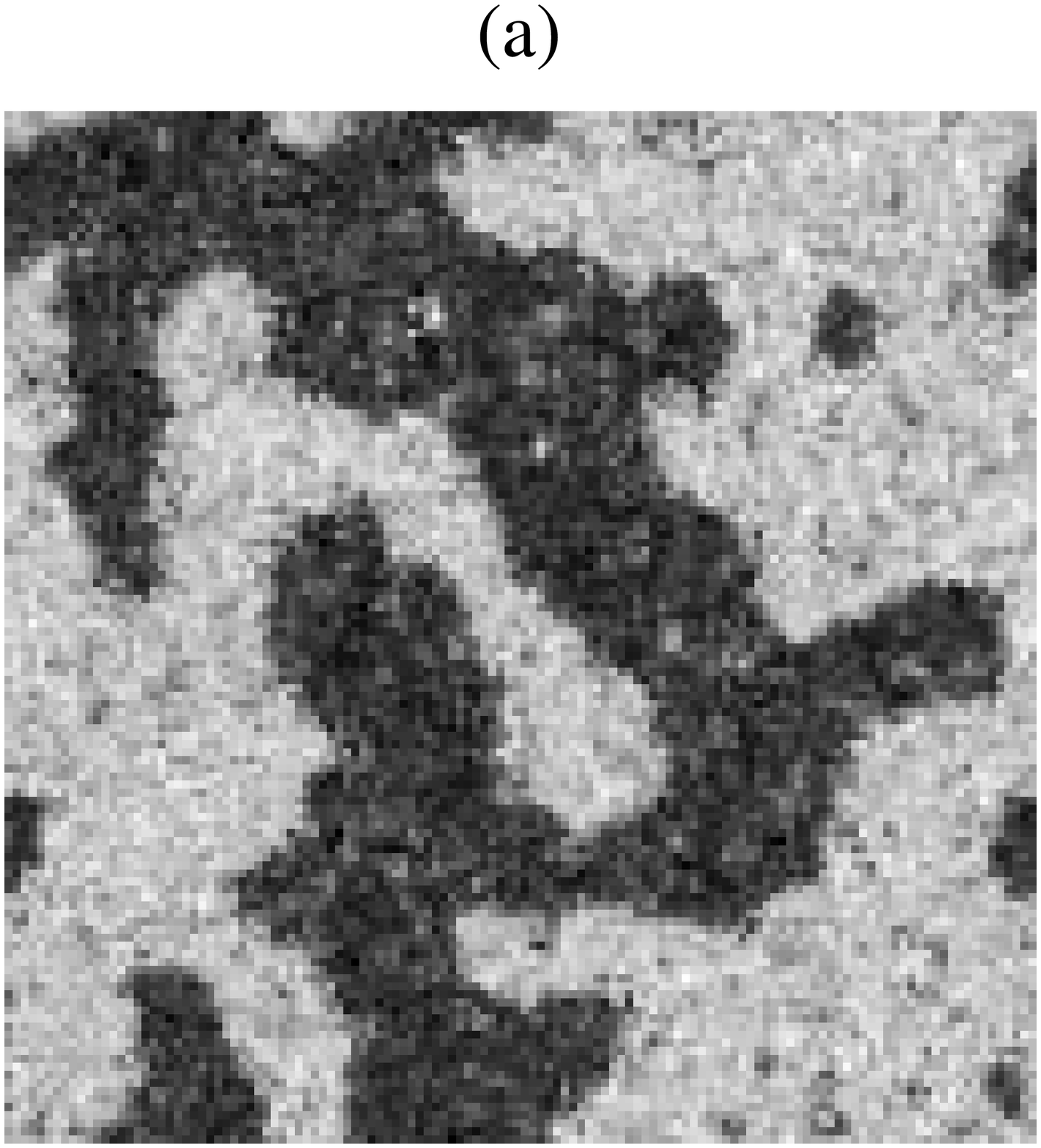}
\epsfxsize=4cm
\epsffile{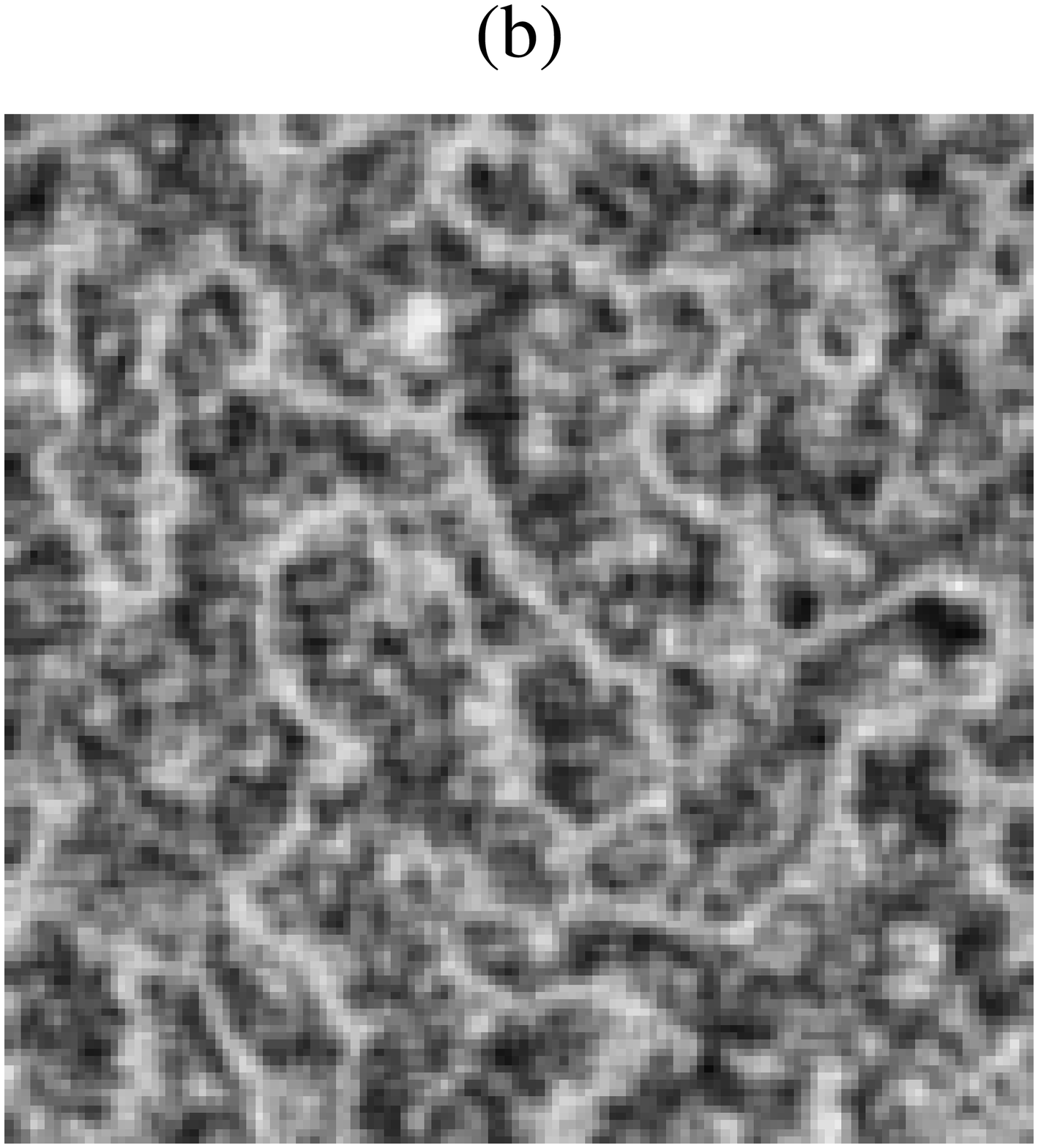}
}

\centerline{
\epsfxsize=4cm
\epsffile{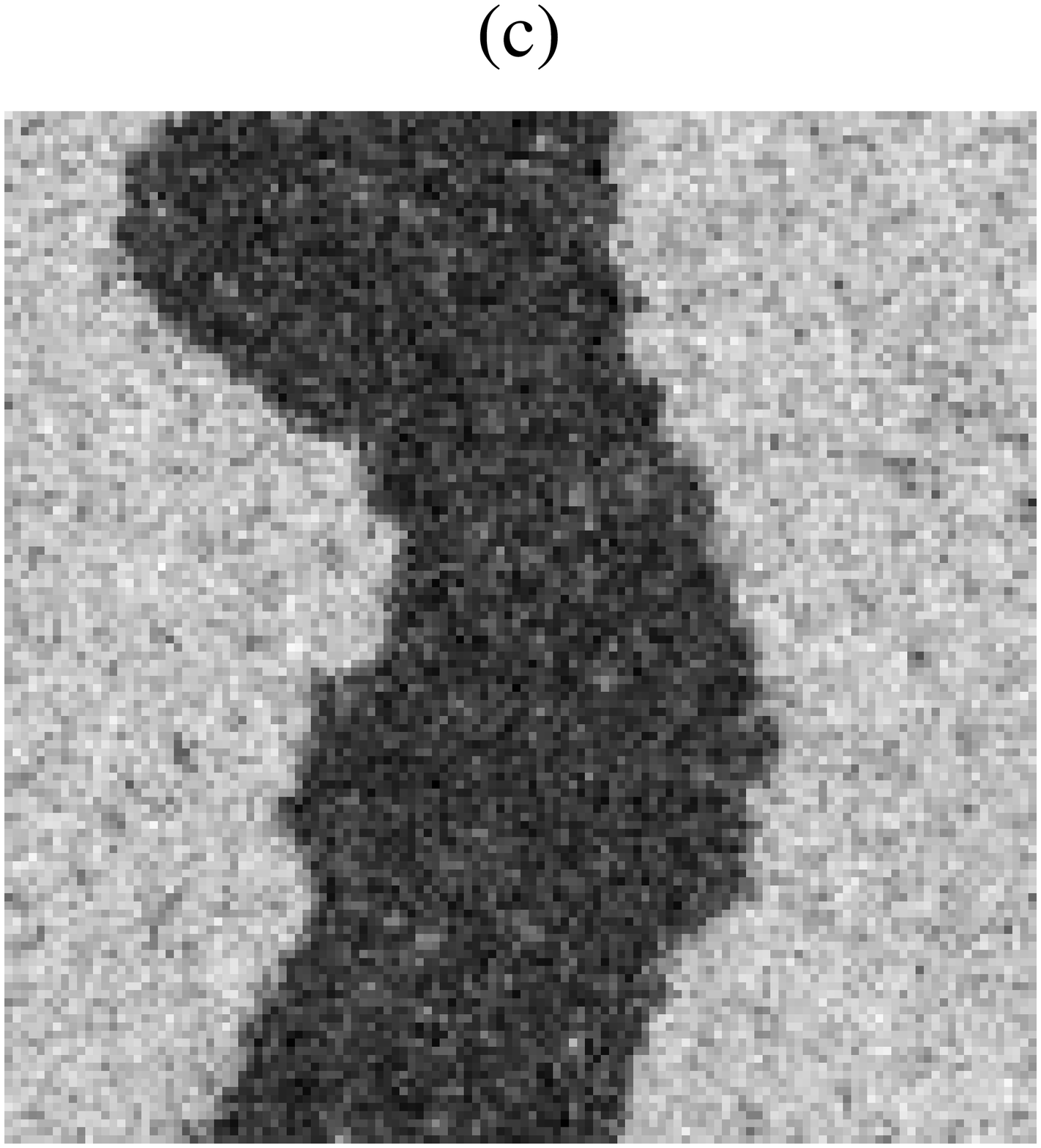}
\epsfxsize=4cm
\epsffile{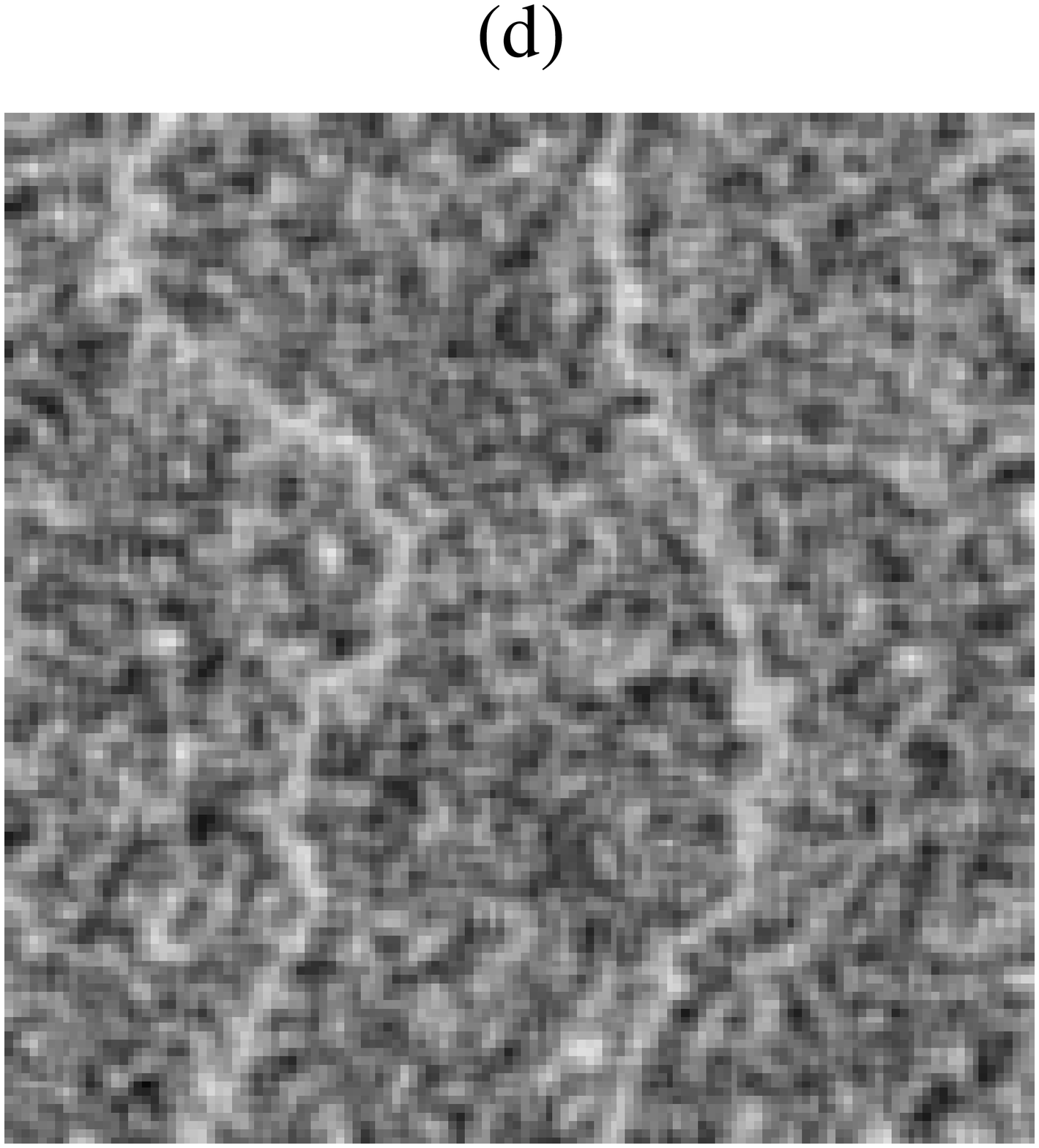}
}

\caption{ Snapshots of (a) order parameter $\phi$ and (b) energy
density $E$ at time $t=15$, (c),(d): id. but at $t=75$.
} 
\label{fig-phi4}
\end{figure}

As stated before, this model is believed to be
in the model~C universality class since the order parameter $\phi$ is coupled
to the conserved energy. \cite{HOHA}
 Indeed, looking at snapshots of the order
parameter and the local energy, defined as:
\begin{equation}
E_i  = \frac12 \dot{\phi_i}^2 - \frac12 m^2 \phi_i^2 + \frac{g}{4!}
\phi_i^4 + \sum_{j=1}^d (\phi_{i+j} - \phi_i)^2  \;,
\end{equation}
one observes that $E_i$ is higher at the interfaces of
ordered domains (Fig.~\ref{fig-phi4}a) as observed in our model~C
around the coexistence concentration $c_s$. 
Given the results described above, the question
of the precise relationship between the microcanonical $\phi^4$
model and model~C now boils down to 
whether we are in the interconnected ($(c_n+c_s)/2<c_i<c_s$) 
or the ordered regime ($c_i>c_s$) of model~C, i.e. 
whether the interfaces (equivalently the disordered domains) 
will thicken or not. 

Our observation that, in fact, $z=2$ \cite{bz-comm} 
indicates that we are in the ordered regime. 
This implies that the conserved field does not undergo
spinodal decomposition. The snapshots of the order parameter and the local
energy taken during coarsening (Fig.~\ref{fig-phi4}) reveal that
the width of the ``antiphase
boundaries'' does not grow. Instead, energy flows into the
bulk of domains in the form of kinetic energy, as observed in \cite{bz-comm}.
The increase of kinetic energy compensates the
energy lost due to the decrease of the total interface length.  
We believe this constitutes further support to our result $z=2$, 
at least for the parameter values studied in \cite{BZ} and \cite{bz-comm}. 
Whether a coexistence regime in the
$\phi^4$-model exists for other parameter values remains,
strictly speaking, an open question. 
In fact, since the system is isolated, 
the total energy and the
``temperature'' cannot be varied independently. 
Upon increasing the energy, the
temperature, which can be identified with the kinetic energy, will also
increase. We therefore only have access to a line in the $(c,T)$-plane of
Fig.~\ref{f1}. (Note that increasing the energy corresponds to decreasing
the concentration.) Investigations of the critical behavior of the
model, where it is a priori important whether one enters the disordered region
from the coexistence region or from the ordered region, are under way
\cite{marcq}.

\subsection{Interfacial properties}

A few years ago, Somoza and Sagui have investigated the effect of
interfacial properties on the morphology of domains in a model~C
system \cite{Somoza}.
They came to the conclusion
that apart from a ``complete wetting'' regime where two ``ordered''
domains ($\psi=\pm 1$) are always separated by a disordered layer
($\psi=0$), there exist also a ``partial wetting'' and even a ``partial
drying'' regime where ordered domains of opposite sign can be in
direct contact.  These regimes were observed by varying the
diffusion constants $K_c$ and $K_{\psi}$ keeping $\Delta x$ and
$\Delta t$ fixed. Thus diffusion of one quantity was enhanced with
respect to the other. 

\begin{figure}
\centerline{
\epsfxsize=4cm
\epsffile{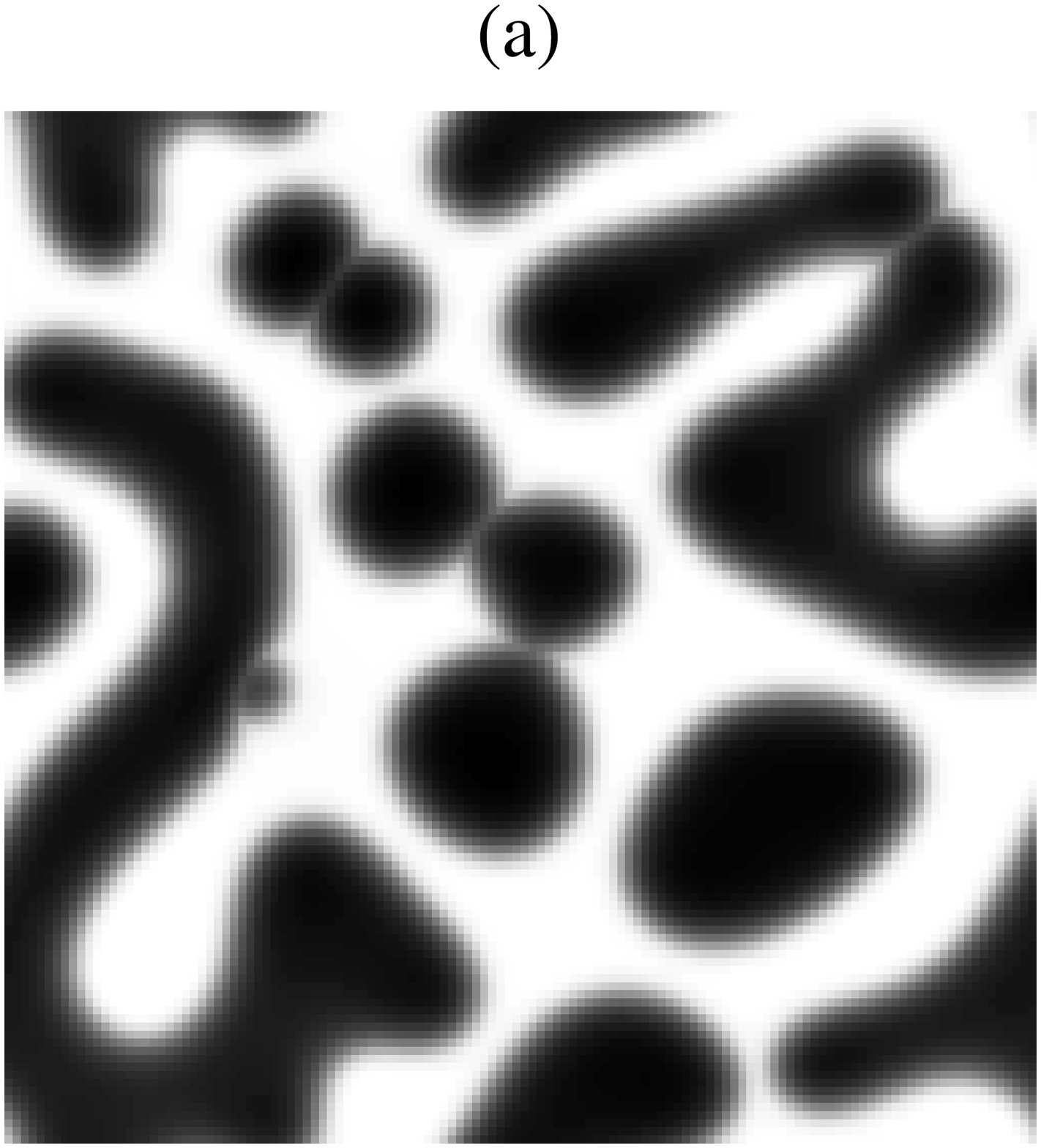}
\epsfxsize=4cm
\epsffile{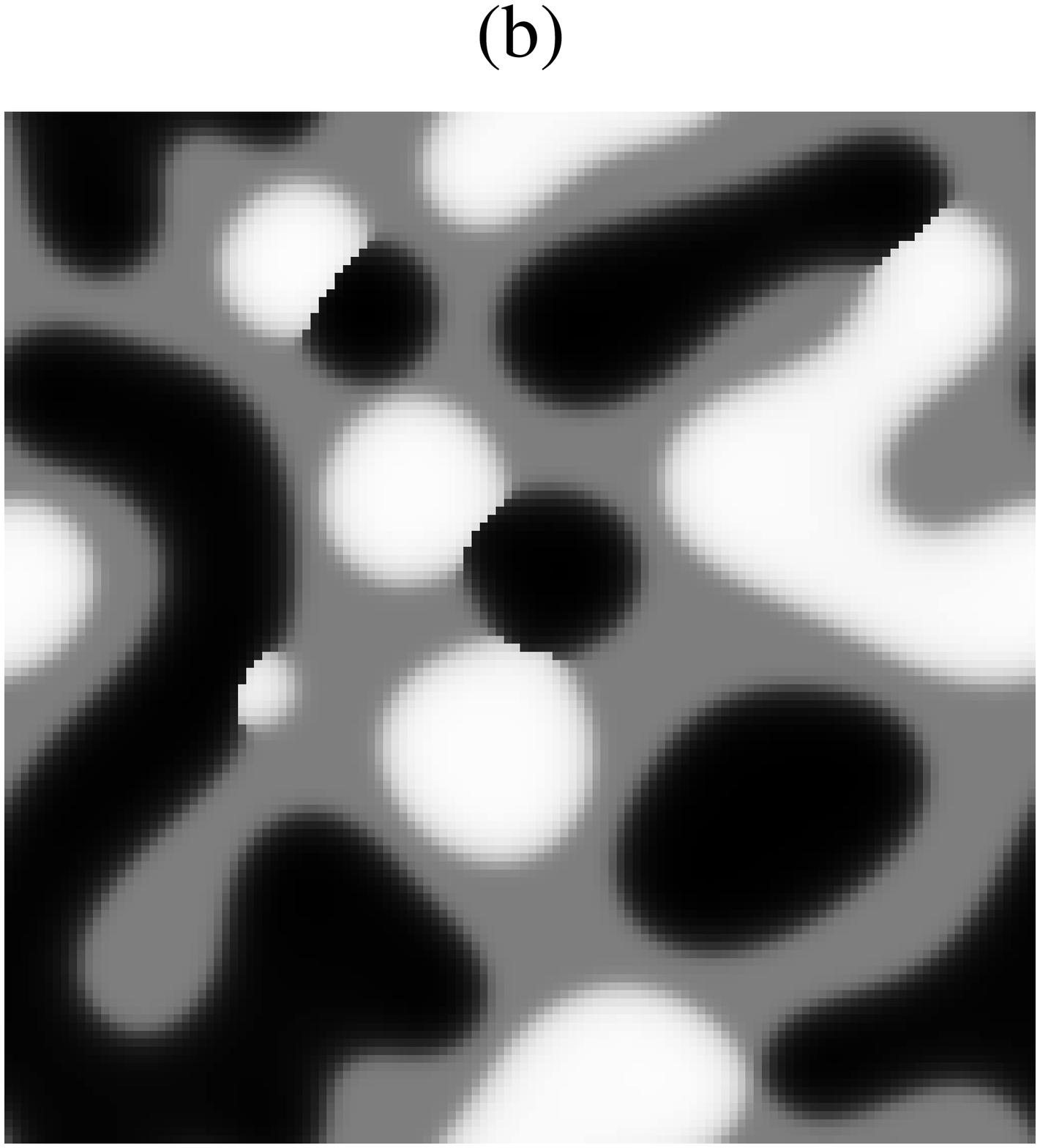}
}

\centerline{
\epsfxsize=4cm
\epsffile{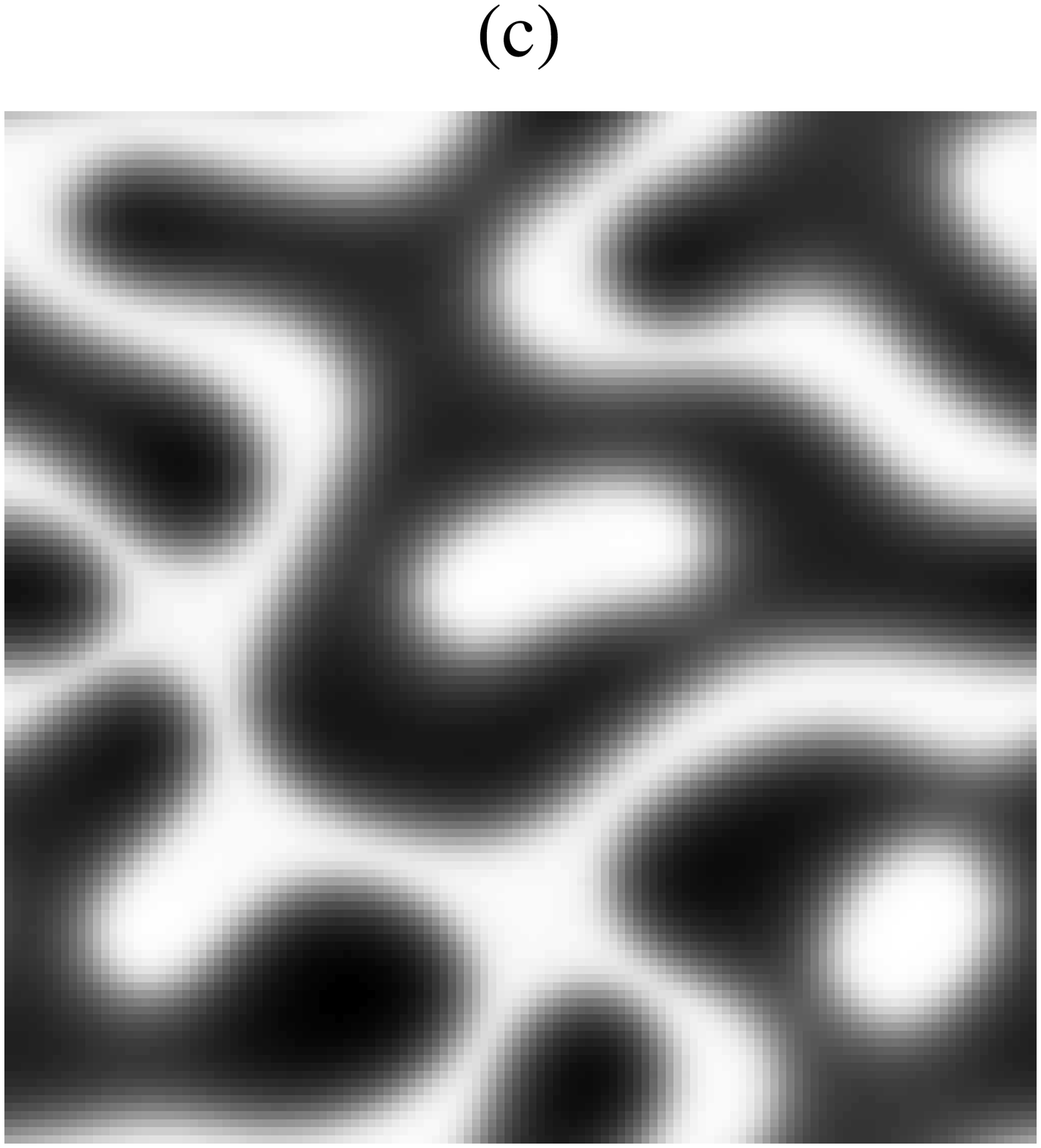}
\epsfxsize=4cm
\epsffile{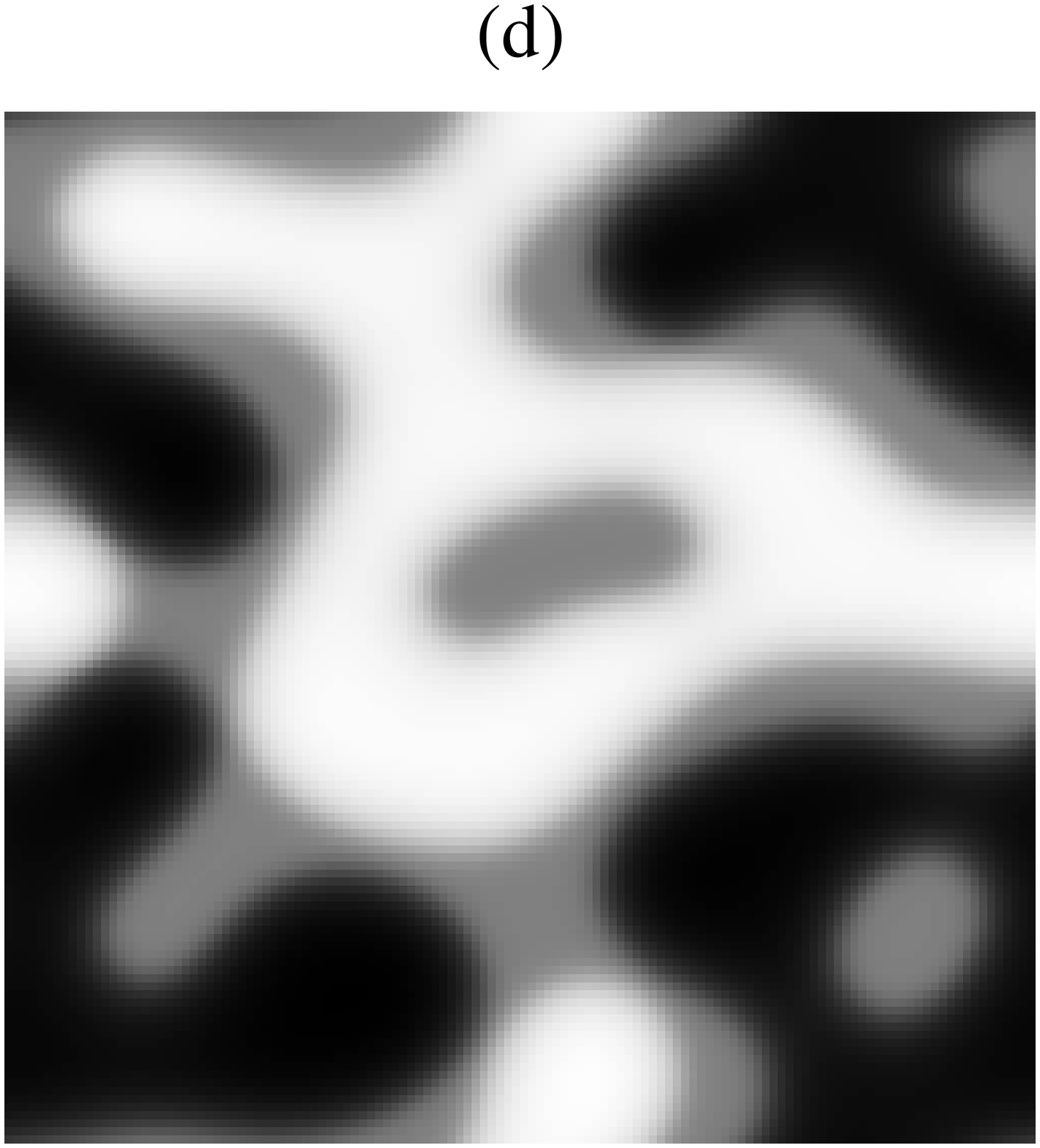}
}
\caption{Snapshots of partial wetting regime $K_{\psi}=0.6$,
$K_{\psi}=1/K_c$ with $\Delta x=1.0$, $\Delta t=0.01$. (a)
Concentration field, (b) Order parameter 
(c),(d), Same system but better resolved: $\Delta =0.3$, $\Delta t=0.005$.   
System size and greyscales are in same as in Fig.~\ref{f4}.} 
\label{f9}
\end{figure}

We first repeated the simulations of Somoza and Sagui using $\Delta x=1$. 
(Fig.~\ref{f9}a). The ordered domains are in direct contact but the
corresponding interfaces are not smooth, an indication that the
equations are not well resolved. 
Next, we performed the same quench but at a higher numerical resolution
($\Delta x=0.3$). As shown in Fig.~\ref{f9}b, no direct contact 
between ordered domains can be observed.
The argument put forward in \cite{Somoza} is that in the partial
drying regime the surface tension is lower between the two ordered phases than 
between the ordered and disordered phases. This surface tension was
measured by calculating the free energy of a numerically obtained
interface profile \cite{andres}. It must be noted, however, that this
procedure is only correct when the obtained profile is the equilibrium
profile. An order-order interface for instance will
desintegrate into two disorder-order interfaces when the surface tension
of the latter is lower. The speed at which this happens depends on the
diffusion constants, since some amount of the conserved quantity has
to diffuse away from the interface. One could thus have the impression
that an order-order interface is stable, when in fact it is very
slowly disintegrating. 


\section{Conclusion}
\label{s5}

We have provided a comprehensive account of domain growth in 
two-dimensional model~C systems, 
unifying and sometimes correcting partial results
present in the literature. Quenches into the coexistence region
($c_n<c_i<c_s$)
all lead to $z=3$ domain growth governed by the conserved field,
independent of the morphology of domains, but subjected to the strength
of initial fluctuations when $c_i<c_1$. 
Quenches into the ordered region are asymptotically dominated by the 
ordering process, leading to $z=2$ growth after a crossover from
slower $z=3$ coarsening.

We also clarified the status of the microcanonical $\phi^4$ model
and showed that in this case sub-critical domain growth correspond to
quenches into the ordered region of model~C.

The picture that thus emerges is that domain growth in model~C systems
below criticality is governed by the growth exponent of either model~A
or model~B. At criticality, on the other hand, one may expect exponents different
from model~A or B. The natural question that then arises is how the
exponents depend on on the initial concentration. 

Another issue open for further investigation is the behavior of the
so-called persistence probability $p$, defined as the probability for a
spin to have remained in its initial phase from $t_0$ up to time
$t$ (see \cite{PERSIS} for a review). 
In the case of model A and B this quantity decays algebraically
with an exponent $\theta$, estimated to be $\theta_{A} \approx 0.21$ \cite{CS}
and $\theta_B \approx 0.25$ \cite{STRA-COM}. The universality
of these exponents is an ongoing debate. 
Measurements of the persistence exponent in model~C, especially in the
interconnected regime where one would expect the model~B value, might help to
settle this issue.  


\end{multicols}

\end{document}